\documentclass[twocolumn,english,aps,pra,superscriptaddress,showpacs,tightenlines]{revtex4-1}
\usepackage{amsmath}
\usepackage{graphicx}
\usepackage{amssymb}
\usepackage{CJK}
\usepackage{color}

\begin{document}

\begin{CJK*}{GBK}{song}

\title{Bound states and the collective dynamics of Distant Quantum Emitters coupled to a chiral waveguide }
\author{Meng Qian \surname{Wu} }
\affiliation{Key Laboratory for Matter Microstructure and Function of Hunan Province, Hunan Research Center of the Basic Discipline for Quantum Effects and Quantum Technologies, Xiangjiang-Laboratory and Department of Physics, Hunan Normal University, Changsha 410081, China}
\author{Ge \surname{Sun}}
\affiliation{Key Laboratory for Matter Microstructure and Function of Hunan Province, Hunan Research Center of the Basic Discipline for Quantum Effects and Quantum Technologies, Xiangjiang-Laboratory and Department of Physics, Hunan Normal University, Changsha 410081, China}
\author{Jing \surname{Lu}}
\thanks{Corresponding author}
\email{lujing@hunnu.edu.cn}
\affiliation{Key Laboratory for Matter Microstructure and Function of Hunan Province, Hunan Research Center of the Basic Discipline for Quantum Effects and Quantum Technologies, Xiangjiang-Laboratory and Department of Physics, Hunan Normal University, Changsha 410081, China}
\author{Lan \surname{Zhou}}
\thanks{Corresponding author}
\email{zhoulan@hunnu.edu.cn}
\affiliation{Key Laboratory for Matter Microstructure and Function of Hunan Province, Hunan Research Center of the Basic Discipline for Quantum Effects and Quantum Technologies, Xiangjiang-Laboratory and Department of Physics, Hunan Normal University, Changsha 410081, China}
\affiliation{Institute of Interdisciplinary Studies, Hunan Normal University, Changsha, 410081, China}

\begin{abstract}
We consider two two-level quantum emitters (QEs) with separations on the order of
the wavelength which are chirally coupled to a one-dimensional (1D) waveguide, and
the electromagnetic field of the 1D waveguide has a direction-dependent velocity,
which produces two field propagation phases on the dynamics of QEs. Their spontaneous
process is examined for QEs having unequal emission rates to the waveguide. It is
found that radiation could be enhanced for both QEs, inhibited for both QEs, enhanced
for one while inhibited for the other, completely suppressed for both QEs. In
particular, the mechanism for radiation completely suppressed is the presence of a
QE-photon bound state.

\end{abstract}
\maketitle

\end{CJK*}\narrowtext

\section{Introduction}

The growing demand for faster and more efficient data transfer and
processing has brought quantum networks to the forefront of research. Nodes
and channels are elementary building blocks for a quantum network, where
quantum information is exchanged in the form of flying qubits interacting
with static qubits. The flying qubits in quantum channels serve to
distribute quantum information. The static qubits in local nodes generate,
process, and route quantum information. Waveguides (they also refers to
optical fibers, microwave transmission lines etc.), have emerged as a new
promising platform for quantum channels since they allow continuous bosonic
modes, the one-dimensional (1D) waveguide is particular interested due to
the strong interaction between quantum emitters (QEs) and propagating
photons. Quantum devices at the single-photon level are proposed for
engineering the transport of photons~\cite{LanPRL101,ZhouOE30,LanPRA78,LiaoPRA92} and
single-photon routing~\cite{lanPRL111,LuPRA89,LuOE23,YangPRA98,routPRL107,routPRA105}. Light which is
tightly transversely confined can exhibit a significant polarization
component along the propagation direction, which breaks the symmetry of
QE-waveguide coupling to the right and left propagating modes. Chiral
interfaces between QEs and waveguides opens the route towards quantum
information processing tasks that cannot be accomplished by the
bidirectional waveguides, for example, target quantum router~\cite%
{routPRA94,WeiPRA89,AhumPRA99}, on-chip circulators\cite%
{127(21)233601,130(23)037001}.

A QE in a 1D waveguide unavoidably decays towards its ground state through
spontaneous emission, but QEs can interact with light in a coherent and
collective way, i.e., single photons emitted by one QE can be reflected or
absorbed and later emitted by other QEs in a 1D waveguide, and the
interference experienced by photons allows the correlations among QEs. This
generation of correlations can be understood in the case of a single
resonant excitation of two QEs, QEs mediated by the exchange of photons
propagating in one dimension are in the superradiant and subradiant states%
\cite{SinhaPRL124,SinhaPRA102,KumlinPRA102}, so an initially factorized atomic state can
spontaneously relax towards a state with finite entanglement\cite%
{PRA94(16)043839}, which is called spontaneous entanglement generation. And
this spontaneous entanglement generation can be enchaned by the vacuum
radiation field of a chiral waveguide \cite {ScI346(14)67,PRA91(15)042116,PRB92(15)155304,PRA101(20)032335,PRR2(20)013369,PRA94(16)012302,PRR2(20)043048,PRL115(15)163603,PRA101(20)013830,PRA95(17)023838}. Many of these studies have focused on the property of the waveguide field: equal propagation velocities in opposite directions. Current platforms allow one to envision direction-dependent velocity. A disparate cooperative decay dynamics of the emitters is found\cite{SinhaPRA107} based on the assumption that two QEs are identical and their coupling strengths to the left-going and right-going modes are equal. In this paper, we consider emission from two QEs located on the axis of a chiral waveguide. In addition to the unequal velocities of photons propagating to the left or to the right. The unequal coupling strengths of the two QEs are also taken into account.

The paper is organized as follows. In Sec.\ref{Sec:2}, we introduce the
Hamiltonian of the two distant QEs coupled to a chiral waveguide, and
present the equation of motion for a single excitation in the system. Then,
in Sec.\ref{Sec:3} we derive the general formulas of the dynamic evolution
with the initial condition in which one atom is in the excited state and
the other in the ground state, and present the condition for the emergence
of the dark state of two QEs. In Sec.\ref{Sec:4}, the local density of photons
emitted by QEs is studied, the trapping of the excitation and the
interference of the localized field modes are found. Then we
make our conclusion in Sec.\ref{Sec:5}

\section{\label{Sec:2}Model and Its Equation of Motion}

\begin{figure}[tbph]
\includegraphics[width=8 cm,clip]{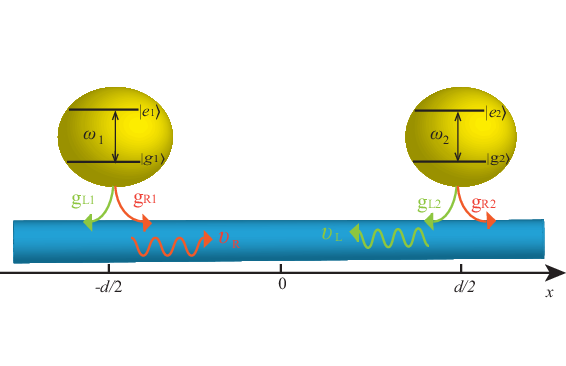}
\caption{A sketch of the system: two-level QEs are located at the positions $%
x=\pm d/2$ along a chiral 1D waveguide and coupled to the left and right
propagating waveguide modes with coupling strengths $g_{Lj}$ and $g_{Rj}$. }
\label{figM}
\end{figure}

The system consists of two two-level quantum emitters with transition
frequency $\omega _{j}$ ($j=1,2$) between the ground $\left\vert
g_{j}\right\rangle $ and excited $\left\vert e_{j}\right\rangle $ states,
and a 1D waveguide in which the photons propagate, as shown in Fig.\ref{figM}%
. The Hamiltonian of the two QEs reads%
\begin{equation}
\hat{H}_{A}=\sum_{j}\omega _{j}\left\vert e_{j}\right\rangle \left\langle
e_{j}\right\vert .  \label{EQ2-01}
\end{equation}%
The 1D waveguide have a continuum of bosonic modes, and the field modes
propagating through the waveguide with unequal velocities $v_{L}$ and $v_{R}$
on the left and the right are denoted by the annihilation operators $\hat{a}%
_{k_{L}}$ and $\hat{a}_{k_{R}}$, respectively. The Hamiltonian of the 1D
waveguide reads
\begin{equation}
\hat{H}_{F}=\sum_{\alpha =R,L}\int_{-\infty }^{\infty }dk_{\alpha }\left(
\omega _{0}+v_{\alpha }k_{\alpha }\right) \hat{a}_{k_{\alpha }}^{\dagger }%
\hat{a}_{k_{\alpha }}  \label{EQ2-02}
\end{equation}%
where $\omega _{0}$ is the central frequency around which a linear
dispersion relation is given by $\omega _{k}=\omega _{0}+v_{R}\left(
k-k_{R}^{0}\right) =\omega _{0}-v_{L}\left( k+k_{L}^{0}\right) $ with $%
\omega _{0}=\omega _{k_{R}^{0}}=\omega _{k_{L}^{0}}$. The integration can
be extended to $\pm \infty $ since weak couplings are considered. The
transitions induced by the photons are described by the Hamiltonian%
\begin{eqnarray}
\hat{H}_{AF} &=&\sum_{j}g_{Lj}\int dk_{L}\hat{a}_{k_{L}}e^{\mathrm{i}\left(
k_{L}-k_{L}^{0}\right) x_{j}}\sigma _{j}^{+}+h.c.  \label{EQ2-03} \\
&&+\sum_{j}g_{Rj}\int dk_{R}\hat{a}_{k_{R}}e^{\mathrm{i}\left(
k_{R}+k_{R}^{0}\right) x_{j}}\sigma_{j}^{+}+h.c.  \notag
\end{eqnarray}%
under the rotating-wave approximation, where $\sigma _{j}^{+}=\left\vert
e_{j}\right\rangle \left\langle g_{j}\right\vert $ is the raising ladder
operator, $g_{\alpha j}=\left\vert g_{\alpha j}\right\vert \exp \left(
\mathrm{i}\varphi _{\alpha j}\right) $ ($\alpha =L,R$) are the coupling
strengths for QE $j$ interacting with a left-going and right-going photon at
the position $x_{j}=\left( -1\right) ^{j}d/2$ and are related to the decay
rate $\gamma _{\alpha j}$ of QE $j$ to the waveguide by $\gamma _{\alpha
j}=2\pi \left\vert g_{\alpha j}\right\vert ^{2}/v_{\alpha }$. For chiral
couplings, we have $\gamma _{Lj}\neq \gamma _{Rj}$. The annihilation and
generation operator of the waveguide satisfy the bosonic commutation
relation $\left[ \hat{a}_{k\alpha },\hat{a}_{k^{\prime }\beta }^{\dag }%
\right] =\delta _{\alpha \beta }\delta \left( k-k^{\prime }\right) $. The
total Hamiltonian that includes the waveguide, the QEs, and their couplings
thus reads $\hat{H}=\hat{H}_{A}+\hat{H}_{F}+\hat{H}_{AF}$.

Since Hamiltonian $\hat{H}$ preserves the total number of excitations, its
ground state $\left\vert \emptyset \right\rangle =\left\vert
g_{1}g_{2}0\right\rangle $ is identical to that of the free part $\hat{H}%
_{A}+\hat{H}_{F}$, i.e., all QEs in the ground state and field in the
vacuum. The states of exciting a single particle from the ground state, $%
\sigma _{j}^{+}\left\vert \emptyset \right\rangle $ and $\hat{a}_{k\alpha
}^{\dagger }\left\vert \emptyset \right\rangle $ are QE-field product
states. The evolution of the system in the single-excitation subspace is
captured by the ansatz%
\begin{equation}
\left\vert \Psi \left( t\right) \right\rangle =\sum_{\alpha }\int dk_{\alpha
}c_{\alpha }\left( k_{\alpha },t\right) \hat{a}_{k_{\alpha }}^{\dagger
}\left\vert \emptyset \right\rangle +\sum_{j}c_{j}\left( t\right) \sigma
_{j}^{+}\left\vert \emptyset \right\rangle  \label{EQ2-04}
\end{equation}%
where $c_{j}\left( t\right) $ and $c_{\alpha }\left( k_{\alpha },t\right) $
are the excitation amplitudes for the QE $j$ and the guided field modes with
wavenumber $k_{\alpha }$, respectively. The Schrodinger equation transforms $%
\left\vert \Psi \left( t\right) \right\rangle $ into the following
differential equations
\begin{subequations}
\label{EQ2-05}
\begin{eqnarray}
\dot{c}_{1}\left( t\right) &=&-\mathrm{i}\omega _{1}c_{1}\left( t\right) -%
\mathrm{i}g_{L1}\int dk_{L}e^{\mathrm{i}\left( k_{L}-k_{L}^{0}\right)
x_{1}}c_{L}\left( k_{L},t\right)  \notag \\
&&-\mathrm{i}g_{R1}\int dk_{R}e^{\mathrm{i}\left( k_{R}+k_{R}^{0}\right)
x_{1}}c_{R}\left( k_{R},t\right) \\
\dot{c}_{2}\left( t\right) &=&-\mathrm{i}\omega _{2}c_{2}\left( t\right) -%
\mathrm{i}g_{L2}\int dk_{L}e^{\mathrm{i}\left( k_{L}-k_{L}^{0}\right)
x_{2}}c_{L}\left( k_{L},t\right)  \notag \\
&&-\mathrm{i}g_{R2}\int dk_{R}e^{\mathrm{i}\left( k_{R}+k_{R}^{0}\right)
x_{2}}c_{R}\left( k_{R},t\right) \\
\dot{c}_{R} &=&-\mathrm{i}g_{R1}^{\ast }e^{-\mathrm{i}\left(
k_{R}+k_{R}^{0}\right) x_{1}}c_{1}\left( t\right) -\mathrm{i}g_{R2}^{\ast
}e^{-\mathrm{i}\left( k_{R}+k_{R}^{0}\right) x_{2}}c_{2}\left( t\right)
\notag \\
&&-\mathrm{i}\left( \omega _{0}+v_{R}k_{R}\right) c_{R}\left( k_{R},t\right)
\\
\dot{c}_{L} &=&-\mathrm{i}g_{L1}^{\ast }e^{-\mathrm{i}\left(
k_{L}-k_{L}^{0}\right) x_{1}}c_{1}\left( t\right) -\mathrm{i}g_{L2}^{\ast
}e^{-\mathrm{i}\left( k_{L}-k_{L}^{0}\right) x_{2}}c_{2}\left( t\right)
\notag \\
&&-\mathrm{i}\left( \omega _{0}-v_{L}k_{L}\right) c_{L}\left( k_{L},t\right)
\end{eqnarray}

\section{\label{Sec:3}emission Dynamics and Bound States}


For the single excitation initial in the QEs, evolution of emitter
excitation amplitudes reads
\end{subequations}
\begin{subequations}
\label{EQ2-06}
\begin{eqnarray}
\dot{C}_{1}\left( t \right) &=&\mathrm{i}\xi _{1}C_{1}\left( t\right) -\beta
_{1}C_{2}\left( t-\tau_{L}\right) \Theta \left( t-\tau_{L}\right) \\
\dot{C}_{2}\left( t\right) &=&\mathrm{i}\xi _{2}C_{2}\left( t\right) -\beta
_{2}C_{1}\left( t-\tau_{R}\right) \Theta \left( t-\tau_{R}\right)
\end{eqnarray}%
by tracing out the field modes and introducing $c_{j}\left( t\right)
=C_{j}\left( t\right) e^{-\mathrm{i}\omega_{e}t}$ with $\omega_{e}
=\left(\omega_{1}+\omega_{2}\right) /2$. Here, $\xi _{j}=\mathrm{i}\frac{\gamma _{j}}{2}+\left( -1\right)
^{j}\delta $, the single-QE decay rate into the waveguide continua is
denoted by $\gamma _{j}=\gamma _{Lj}+\gamma _{Rj}$, $\delta =\left(
\omega_{1}-\omega_{2}\right) /2$ is the detuning of the QEs. $\Theta
\left( t\right) $ is the Heaviside-step function, $\tau _{\alpha }=d/v_{\alpha }$ is the
time delay due to the traveling time of a photon exchanged between QEs. The
parameter $\beta _{1}=\sqrt{\gamma _{L1}\gamma _{L2}}e^{\mathrm{i}\left(
\theta _{L}+\varphi _{L}\right) }$ denotes the strength of interaction
mediated by photons propagating to the left from QE 2 to 1, while $\beta
_{2}=\sqrt{\gamma _{R1}\gamma _{R2}}e^{\mathrm{i}\left( \theta _{R}-\varphi
_{R}\right) }$ corresponds to photons propagating to the right from QE 1 to
2, where $\theta _{\alpha }=\left( v_{\alpha }k_{\alpha
}^{0}+\omega_{e}-\omega_{0}\right) \tau _{\alpha }$ is the propagation phase
acquired by the resonant photon in the waveguide, a phase $\varphi _{\alpha
}=\varphi _{\alpha 1}-\varphi _{\alpha 2}$ is also acquired at the coupling
points.
\begin{figure}[tbp]
\includegraphics[width=9cm, height=8cm]{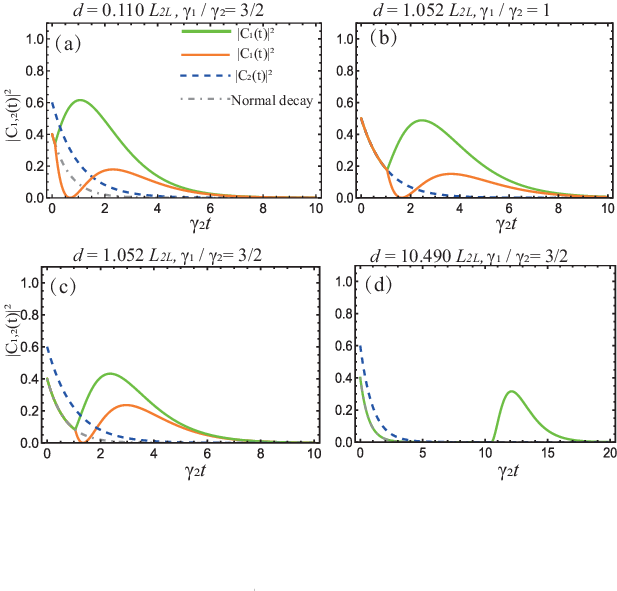}
\caption{The population dynamics $|C_{j}(t)|^{2}$ for QEs with initial
condition $C_1(0)=\protect\sqrt{\protect\gamma_2/(\protect\gamma_2+\protect%
\gamma_1)}, C_2(0)=i\protect\sqrt{\protect\gamma_1/(\protect\gamma_2+\protect%
\gamma_1)}$ for a) $v_L / c = 0.909$, $v_R / c = 0.740$; (b,c) $v_L / c =
0.950$, $v_R / c = 0.789$; (d) $v_L / c = 0.953$, $v_R / c = 0.795$, the
remaining parameters $\protect\delta = 0$, $\protect\omega_e = 500\protect%
\gamma_2$, $\protect\theta_L = (n + \frac{1}{2})\protect\pi$, $\protect\theta%
_R = (m + \frac{1}{2})\protect\pi$, $\protect\varphi _{R}=0$ $\protect\varphi%
_L = 2\protect\pi$ (solid green curve) and $\protect\varphi_L = \protect\pi$
(solid orange curve).}
\label{fig1}
\end{figure}

The delay differential equations (\ref{EQ2-06}) describe the mutual
influence of the QEs located at a distance by sharing the same
electromagnetic modes. Its Laplace transform leads to
\end{subequations}
\begin{subequations}
\label{EQ3-01}
\begin{align}
C_{1}\left( s\right) & =\frac{\left( s-\mathrm{i}\xi _{2}\right) C_{1}\left(
0\right) -\beta _{1}e^{-s\tau _{L}}C_{2}\left( 0\right) }{\left( s-\mathrm{i}%
\xi _{1}\right) \left( s-\mathrm{i}\xi _{2}\right) -\beta _{1}\beta
_{2}e^{-s\left( \tau _{L}+\tau _{R}\right) }}, \\
C_{2}\left( s\right) & =\frac{\left( s-\mathrm{i}\xi _{1}\right) C_{2}\left(
0\right) -\beta _{2}e^{-s\tau _{R}}C_{1}\left( 0\right) }{\left( s-\mathrm{i}%
\xi _{1}\right) \left( s-\mathrm{i}\xi _{2}\right) -\beta _{1}\beta
_{2}e^{-s\left( \tau _{L}+\tau _{R}\right) }}.
\end{align}%
Defining $t_{n}=t-n\left( \tau _{L}+\tau _{R}\right) $, the inverse Laplace
transform yields the time dependent amplitudes of the QEs
\end{subequations}
\begin{subequations}
\label{EQ3-02}
\begin{eqnarray}
C_{1}(t) &=&\sum_{n}C_{1}\left( 0\right) \left[ B_{n}\left( \xi _{1}\right)
+A_{n}\left( \xi _{2}\right) \right] \Theta \left( t_{n}\right) \\
&&-\sum_{n}C_{2}\left( 0\right) \left[ D_{n}\left( \xi _{1}\right)
+D_{n}\left( \xi _{2}\right) \right] \Theta \left( t_{n}-\tau _{L}\right)
\notag \\
C_{2}(t) &=&\sum_{n}C_{2}\left( 0\right) \left[ B_{n}\left( \xi _{2}\right)
+A_{n}\left( \xi _{1}\right) \right] \Theta \left( t_{n}\right) \\
&&-\sum_{n}C_{1}\left( 0\right) \left[ E_{n}\left( \xi _{1}\right)
+E_{n}\left( \xi _{2}\right) \right] \Theta \left( t_{n}-\tau _{R}\right)
\notag
\end{eqnarray}%
where the functions are defined as
\end{subequations}
\begin{subequations}
\label{EQ3-03}
\begin{eqnarray}
A_{n}\left( \xi _{j}\right) &=&\lim_{z\rightarrow \xi _{j}}\frac{d^{n-1}}{%
dz^{n-1}}\frac{\left( -1\right) ^{n}\beta _{1}^{n}\beta _{2}^{n}e^{\mathrm{i}%
zt_{n}}}{\left( n-1\right) !\left( z-\xi _{j^{\prime }}\right) ^{n+1}} \\
B_{n}\left( \xi _{j}\right) &=&\lim_{z\rightarrow \xi _{j}}\frac{d^{n}}{%
dz^{n}}\frac{\left( -1\right) ^{n}\beta _{1}^{n}\beta _{2}^{n}e^{\mathrm{i}%
zt_{n}}}{n!\left( z-\xi _{j^{\prime }}\right) ^{n}} \\
D_{n}\left( \xi _{j}\right) &=&\lim_{z\rightarrow \xi _{j}}\frac{d^{n}}{%
dz^{n}}\frac{-\mathrm{i}\left( -1\right) ^{n}\beta _{1}^{n+1}\beta
_{2}^{n}e^{\mathrm{i}z\left( t_{n}-\tau _{L}\right) }}{n!\left( z-\xi
_{j^{\prime }}\right) ^{n+1}} \\
E_{n}\left( \xi _{j}\right) &=&\lim_{z\rightarrow \xi _{j}}\frac{d^{n}}{%
dz^{n}}\frac{-\mathrm{i}\left( -1\right) ^{n}\beta _{1}^{n}\beta
_{2}^{n+1}e^{\mathrm{i}z\left( t_{n}-\tau _{R}\right) }}{n!\left( z-\xi
_{j^{\prime }}\right) ^{n+1}}
\end{eqnarray}%
with subscripts $j\neq j^{\prime }\in \left\{ 1,2\right\} $. In the
following cases: 1) $\min_{\alpha }\tau _{\alpha }\rightarrow \infty $; 2) $%
\gamma _{L1}=0$ ($\gamma _{L2}=0$) and $\gamma _{R2}=0$ ($\gamma _{R1}=0$),
each QE behaves independently and QE $j$ decays exponentially with rate $%
\gamma _{j}$ to its ground state accompanied by an irreversible release of
energy to the vacuum of a waveguide. In the case of $\gamma _{Lj}=0$ ($%
\gamma _{Rj}=0$), QE 1 (QE 2) decays exponentially with rate $\gamma _{1}$ ($%
\gamma _{2}$) all the time since QE 1 (QE 2) is coupled only to
right-propagating (left-propagating) modes and thus is not able to interact
with its partner, however, the decay behavior of QE 1 (QE 2) is different
before and after time $\tau _{R}(\tau _{L})$. At first, QE 2 (QE 1) also
decays exponentially with rate $\gamma _{2}$ ($\gamma _{1}$). As time
increases until $t\geq \tau _{R}(\tau _{L})$, the emitted photon propagates
along the waveguide will be absorbed by QE 2 (1), and later QE 2 (1) re-emit
the photon to the right (left) again, which propagate away from the QEs. In
Fig.~\ref{fig1}, we plot the excitation probability as a function of time
when QEs are initial in $\left\vert \Psi (0)\right\rangle =\sqrt{\frac{%
\gamma _{2}}{\gamma _{2}+\gamma _{1}}}\left( \left\vert
e_{1}g_{2}\right\rangle -\mathrm{i}\sqrt{\frac{\gamma _{1}}{\gamma _{2}}}%
\left\vert g_{1}e_{2}\right\rangle \right) $ at the condition $\delta =0$
and $\gamma _{j}=\gamma _{Lj}$. The QE's excitation is determined by the
ratio of the distance $d$ to the characteristic wavelength $L_{j\alpha
}\equiv v_{\alpha }/\gamma _{\alpha j}$. There are three different regimes:
QEs close to each other characterized by $d\ll L_{2L}$ (see Fig.~\ref{fig1}%
a), the distance between the QEs comparable to the coherence length $d\sim
L_{2L}$ ( see Fig.~\ref{fig1}b,c) and the interatomic distance much larger
than the coherent length $d\gg L_{2L}$ (see Fig.~\ref{fig1}d). It can be
found from Fig.~\ref{fig1} that QE 1 evolves as if QE 2 were absent since QE
1 cannot radiate in the right-propagating mode, the evolution of QE 1 at time
$t$ depends on the state of QE 2 at the retarded time $t-\tau _{L}$. The
decay of QE 1 can be inhibited or enhanced by changing the phases of the
coupling strengths for fixed velocities after the field from one QE reaches
the other (see Fig.~\ref{fig1}a-c), so does it by adjusting the velocities
for fixed coupling strengths since $v_\alpha$ effect the phase $\theta_\alpha$.

\begin{figure}[tbp]
\includegraphics[width=9cm, height=8.5cm]{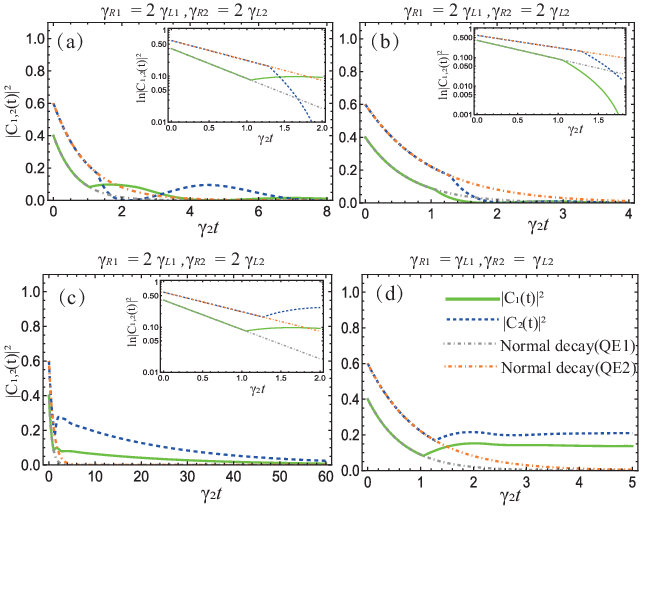}
\caption{The population dynamics $|C_{j}(t)|^{2}$ for QEs with initial
condition $C_1(0)=\protect\sqrt{\protect\gamma_2/(\protect\gamma_2+\protect%
\gamma_1)}, C_2(0)=i\protect\sqrt{\protect\gamma_1/(\protect\gamma_2+\protect%
\gamma_1)}$ in the presence of delay $d=1.0526 L_{2L}$. Parameters are set
as follow: $\protect\delta =0$, $\protect\omega _{e}=500\protect\gamma _{2}$%
, $\protect\gamma _{1}=3\protect\gamma _{2}/2$, $\protect\varphi _{R}=0$, $%
v_{L}/c=0.950\ $and $\protect\theta _{L}=(n+\frac{1}{2})\protect\pi $: (a) $%
\protect\theta _{R}=(m+\frac{1}{2})\protect\pi $,$v_{R}/c=0.774$,$\protect%
\varphi _{L}=2\protect\pi- \protect\varphi _{R}$; (b) $\protect\theta %
_{R}=(m+\frac{1}{2})\protect\pi $,$v_{R}/c=0.774$,$\protect\varphi _{L}=%
\protect\pi $; (c,d) $\protect\theta _{R}=(2m+\frac{1}{2})\protect\pi $,$%
v_{R}/c=0.785$,$\protect\varphi _{L}=2\protect\pi $.}
\label{fig2}
\end{figure}

Generally, a photon emitted by one QE into the waveguide will propagate to
the left and right, and some may be reabsorbed by the other until time $\tau
_{R}$ or $\tau _{L}$, then the photon is emitted again by the other QE, The
process of absorption and emission would be repeated as time increases, the
multiple absorption and emission of a photon generate the correlation
between two QEs, which also modified the emission rate from QEs compared to
an independent emission~\cite{mod-rate1,mod-rate2,RMP95(23)015002}. To have
a clear view of coherent interactions between two QEs, we will assume that
two QEs have equal transition frequencies, i.e., $\delta =0$, and plot the
population $|C_{j}(t)|^{2}$ as a function of time in Fig.~\ref{fig2} with
the same initial state of QEs to Fig.~\ref{fig1} in the presence of delay
with $d\sim \max L_{j\alpha }$. All QEs initially decay exponentially,
after the emitted photon encounters the other QE, the probability of QEs
no longer follow an exponential decay, as indicated by the inset showing
the time evolution on a logarithmic scale. Fig.~\ref{fig2}a has a revival
after the exponential decay and then decays again. The interference in
the propagating modes can enhance the decay of both QEs as shown in
Fig.~\ref{fig2}b, inhibit the decay of both QEs as shown in Fig.~\ref{fig2}c,
accelerate the decay of one QE while slowing the decay of the other as shown
in Fig.~\ref{fig2}a, or completely suppress the emission of photons into
the waveguide in the stationary regime ($t\rightarrow \infty $) as shown
in Fig.~\ref{fig2}d. To understand the completely suppression of emission,
we find the pure imaginary pole $s=0$ of Eq.(\ref{EQ3-01}) under the condition
\end{subequations}
\begin{subequations}
\label{EQ3-04}
\begin{eqnarray}
2p\pi &=&\theta _{L}+\theta _{R}+\varphi _{L}-\varphi _{R}, \\
\frac{\gamma _{1}\gamma _{2}}{4} &=&\sqrt{\gamma _{L1}\gamma _{L2}\gamma
_{R1}\gamma _{R2}}.
\end{eqnarray}%
For the parameters with $\gamma _{Lj}=\gamma _{Rj}=\gamma _{j}/2$, we obtain
\end{subequations}
\begin{subequations}
\label{EQ3-05}
\begin{eqnarray}
C_{1s}&=&\frac{\gamma _{2}C_{1}\left( 0\right) -\sqrt{\gamma
_{1}\gamma _{2}}e^{\mathrm{i}\left( \theta _{L}+\varphi _{L}\right)
}C_{2}\left( 0\right) }{\left( \gamma _{2}+\gamma _{1}\right) +\left( \tau
_{L}+\tau _{R}\right)\gamma _{1}\gamma _{2}/2} \\
C_{2s}&=& -\sqrt{\frac{\gamma _{1}}{\gamma _{2}}}e^{-\mathrm{i}\left( \theta _{L}+\varphi _{L}\right)}C_{1s}
\end{eqnarray}
in the stationary regime using the final value theorem, i.e., after all
unstable states die out, state
\end{subequations}
\begin{equation}
\label{EQ3-06}
\left\vert d\right\rangle =\sqrt{\frac{\gamma _{2}}{\gamma _{1}+\gamma _{2}}}%
\left\vert eg\right\rangle -\sqrt{\frac{\gamma _{1}}{\gamma _{1}+\gamma _{2}}%
}e^{-\mathrm{i}\left( \theta _{L}+\varphi_{L} \right) }\left\vert ge\right\rangle
\end{equation}%
traps an amount of excitation in the QEs and leads to the spontaneous
generation of the entanglement between distant QEs.


\section{\label{Sec:4}emitted photonic modes}

\begin{figure}[tbp]
\includegraphics[width=10cm, height=8cm]{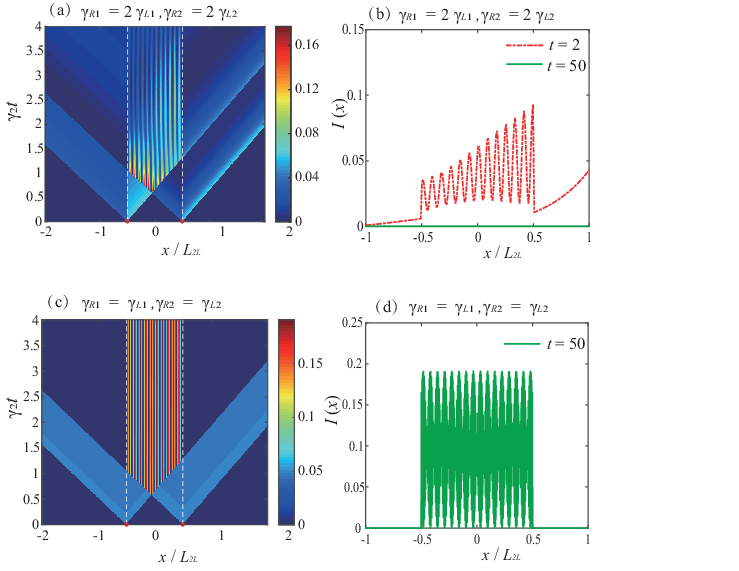}
\caption{The local photon density  $I(x,t)$ for QEs with initial condition $C_1(0)=\protect\sqrt{\protect\gamma_2/(\protect\gamma_2+\protect%
\gamma_1)}, C_2(0)=i\protect\sqrt{\protect\gamma_1/(\protect\gamma_2+\protect%
\gamma_1)}$ in the presence of delay $d = 1.0526 L_{2L}$. With parameters : $\protect\delta =0$, $\protect\omega _{e}=500\protect\gamma _{2}$%
, $\protect\gamma _{1}=3\protect\gamma _{2}/2$, $\protect\varphi _{R}=0$, $%
v_{L}/c=0.950\ $and $\protect\theta _{L}=(n+\frac{1}{2})\protect\pi $,$\protect\varphi _{L}=%
\protect 2\pi $: (a,b) $%
\protect\theta _{R}=(m+\frac{1}{2})\protect\pi $,$v_{R}/c=0.774$,$\protect%
\varphi _{L}=2\protect\pi$;(c,d) $\protect\theta %
_{R}=(2m+\frac{1}{2})\protect\pi $,$v_{R}/c=0.785$.}
\label{fig3}
\end{figure}
Since the probability of ending up in state $\left\vert d\right\rangle $ is
not unit, the remaining amount of excitation is in the waveguide. Studying
the dynamics of the field emitted by the non-Markovian behavior of the QEs
helps us to get more physical insight in the steady state of the
waveguide-QED system. We consider the local photon density $I\left(
x,t\right) =\left\langle \Psi \left( t\right) \right\vert \hat{a}^{\dagger
}\left( x\right) \hat{a}\left( x\right) \left\vert \Psi \left( t\right)
\right\rangle $ at position $x$ and time $t$~\cite{GePRA111(25),HypAQT08(25)}%
, where real-space field annihilation operator at the point $x$ of the
waveguide can be expressed as%
\begin{equation}
\hat{a}\left( x\right) =\sum_{\alpha
=R,L}\sqrt{\frac{v_\alpha}{2\pi}}\int_{-\infty }^{+\infty }e^{\mathrm{i}k_{\alpha }x}\hat{a}_{k_{\alpha
}}dk_{\alpha }.
\end{equation}%
In the one excitation subspace, the local photon density $I\left( x,t\right)
=\left\vert \psi \left( x,t\right) \right\vert ^{2}\equiv \left\vert
\left\langle 0\right\vert \hat{a}\left( x\right) \left\vert \Psi \left(
t\right) \right\rangle \right\vert ^{2}$. With the state $\left\vert \Psi
\left( t\right) \right\rangle $ given in Eq.(\ref{EQ2-04}), the real-space
field amplitude reads%
\begin{align*}
\mathrm{i}\psi(x,t) =&\, \sqrt{\gamma_{L1}} e^{\mathrm{i}f_{L1}} C_1\left(t_L + \frac{\tau_L}{2}\right) \theta\left(t_L + \frac{\tau_L}{2}\right) \theta(x_1 - x) \\
&+ \sqrt{\gamma_{L2}} e^{\mathrm{i}f_{L2}} C_2\left(t_L - \frac{\tau_L}{2}\right) \theta\left(t_L - \frac{\tau_L}{2}\right) \theta(x_2 - x) \\
&+ \sqrt{\gamma_{R1}} e^{\mathrm{i}f_{R1}} C_1\left(t_R - \frac{\tau_R}{2}\right) \theta\left(t_R - \frac{\tau_R}{2}\right) \theta(x - x_1) \\
&+ \sqrt{\gamma_{R2}} e^{\mathrm{i}f_{R2}} C_2\left(t_R + \frac{\tau_R}{2}\right) \theta\left(t_R + \frac{\tau_R}{2}\right) \theta(x - x_2)
\\[1ex]
\tag{\theequation} \stepcounter{equation}
\end{align*}
where we have defined $t_{L}=t+x/v_{L}$, $t_{R}=t-x/v_{R}$ and the phases
\begin{eqnarray*}
f_{L1} &=&-\omega _{e}\left( t+\frac{x}{v_{L}}\right) +\omega _{0}\frac{x}{%
v_{L}}-\frac{\theta _{L}}{2}-\varphi _{L1}, \\
f_{L2} &=&-\omega _{e}\left( t+\frac{x}{v_{L}}\right) +\omega _{0}\frac{x}{%
v_{L}}+\frac{\theta _{L}}{2}-\varphi _{L2}, \\
f_{R1} &=&-\omega _{e}\left( t-\frac{x}{v_{R}}\right) -\omega _{0}\frac{x}{%
v_{R}}+\frac{\theta _{R}}{2}-\varphi _{R1}, \\
f_{R2} &=&-\omega _{e}\left( t-\frac{x}{v_{R}}\right) -\omega _{0}\frac{x}{%
v_{R}}-\frac{\theta _{R}}{2}-\varphi _{R2}.
\end{eqnarray*}%
Figure~\ref{fig3} numerically shows the dependence of the local photon density $I\left(
x,t\right) $ on time and coordinator for the distance between the QEs
comparable to the coherence length $L_{2L}$. As the wave radiated by QEs, it first
propagates away from the QEs, then some wave propagates back and forth between
the regime sandwiched by the two QEs while its amplitude is damped due to energy
exchanged between the QEs and the waveguide. The back and forth waves superimpose
to produce a series of alternating bright and dark fringes. As time increases, the
local density in space at sufficient long time vanishes (see the green line in
Fig.~\ref{fig3}b), however, Fig.~\ref{fig3}(c) shows steady fringes at sufficient
long time, one can observe an oscillating wave fixed in space in Fig.~\ref{fig3}(d),
whose wavenumber $(\omega_e-\omega_0)(v_L^{-1}+v_R^{-1})$, this indicates that
the field comes to a time-independent steady state, i.e., single photons are
localized in a finite regime. This emerge due to the parameters in Fig.~\ref{fig3}(c,d)
satisfying the condition in Eq.~(\ref{EQ3-04}).  This steady state of the field is
a localized eigenmode with energy eigenvalue residing directly in the scattering
continuum, which is called bound state in the continuum (BIC). The superposition
of QEs's dark state and the BIC of singe photons forms a QE-photon bound state
of the waveguide quantum electrodynamics system we studied.


\section{\label{Sec:5}Conclusion}

In this paper, we consider two distant QEs chirally coupled to a 1D
waveguide and study the emission dynamics of a single coherent excitation
coherently shared by two QEs. In the regime that the distance between
two QEs is comparable to tthe coherent length of a spontaneously emitted
photon. The interference of the multiple emission and absorption of photons
produces the following emission dynamics: enhance the decay of both QEs; inhibit
the decay of both QEs; accelerate the decay of one QE while slowing the
decay of the other; completely suppress the emission of photons into the
waveguide, leading to a dark state of two QEs. Via analyzing the local photon
density, we found that single photons can be trapped in the regime sandwiched
by two QEs for the same parameters to the dark state, the trapping energy
resides directly in the scattering continuum, i.e., a BIC is formed. We note
that the BIC formed by different QE-waveguide coupling strengths has not been
found before. The superposition of dark state and the BIC of singe photons
forms a QE-photon bound state.

\begin{acknowledgments}
This work was supported by NSFC Grants No.12247105, No.92365209, No.12421005, XJ-Lab Key Project (23XJ02001),
and the Science $\And $ Technology Department of Hunan Provincial Program (2023ZJ1010).
\end{acknowledgments}

\end{document}